\documentclass[conference]{IEEEtran}
\IEEEoverridecommandlockouts
\usepackage{cite}
\usepackage{amsmath,amssymb,amsfonts}
\usepackage{algorithmic}
\usepackage{algorithm}
\usepackage{graphicx}
\usepackage{textcomp}
\usepackage{xcolor}
\usepackage{multirow}
\usepackage{adjustbox}
\usepackage{orcidlink}
\usepackage{float}
\usepackage{subfigure}
\usepackage{makecell}
\usepackage{subcaption}
\usepackage{braket}
\usepackage{enumitem}
\usepackage{quantikz}
\usepackage{adjustbox}

\usepackage{tikz}
\usepackage{quantikz}
\usepackage{amsmath}

\newcommand{\rowSep}{0.25cm}
\newcommand{\rowSepT}{0.4cm}
\newcommand{\colSep}{0.2cm}
\newcommand{\colSepT}{0.16cm}
\tikzset{slice/.append style={draw=blue}}

\newcommand{\attackCNOTtwo}{
	\begin{quantikz} [row sep= \rowSepT, column sep= 0.215cm, transparent]
		\lstick{$\ket{1}$}  & \ctrl{1} 	& \qw & \qw	\slice{n=1}   & \qw & \qw & \ctrl{1} 	& \qw & \qw	\slice{n=2}   & \qw & \qw & \ctrl{1} 	& \qw & \qw	\slice{n=3}   & \qw & \qw & \ctrl{1} 	& \qw & \qw	\slice{n=4} 	& \qw & \qw & \ctrl{1} 	& \qw & \qw	\slice{n=5} 	& \qw & \rstick{} \\
		\lstick{$\ket{x}$}  & \targ{ } 	& \qw & \qw	              & \qw & \qw & \targ{ } 	& \qw & \qw	              & \qw & \qw & \targ{ } 	& \qw & \qw	              & \qw & \qw & \targ{ } 	& \qw & \qw	            	& \qw & \qw & \targ{ } 	& \qw & \qw	            	& \qw & \rstick{}
	\end{quantikz}
}

\newcommand{\attackAltCNOT}{
	\begin{quantikz} [row sep= \rowSepT, column sep= \colSep, transparent]
		\lstick{$\ket{1}$} & \ctrl{1} 	& \targ{  } 	& \targ{} 	& \ctrl{1} 	& \qw	\slice{n=1}   & \qw & \qw{} 	& \targ{  }		& \qw	\slice{n=2}   & \qw & \targ{} 	& \ctrl{1} 	& \qw	\slice{n=3}   & \qw & \qw{} 	& \targ{  }  	& \qw	\slice{n=4}   & \qw & \rstick{} \\
		\lstick{$\ket{x}$} & \targ{ } 	& \ctrl{-1} 	& \qw    	& \targ{ } 	& \qw	              & \qw & \targ{} 	& \ctrl{-1}		& \qw	              & \qw & \qw    	& \targ{ } 	& \qw	              & \qw & \targ{} 	& \ctrl{-1} 	& \qw	              & \qw & \rstick{}
	\end{quantikz}
}

\newcommand{\attackSACtwo}{
	\begin{quantikz} [row sep= \rowSep, column sep= \colSep, transparent]
		\lstick{$\ket{1}$} & \qw  & \gate{H} 	& \ctrl{1} 	& \targ{  } 	& \qw	\slice{n=1}   & \qw & \ctrl{1} 	& \targ{  } & \qw	\slice{n=2}   & \qw & \ctrl{1} 	& \targ{  } 	& \qw	\slice{n=3}   & \qw & \ctrl{1} 	& \targ{  } & \qw	\slice{n=4} & \qw 	 & \rstick{} \\
		\lstick{$\ket{x}$} & \qw  & \gate{H} 	& \targ{ } 	& \ctrl{-1} 	& \qw	              & \qw & \targ{ } 	& \ctrl{-1} & \qw	              & \qw & \targ{ } 	& \ctrl{-1} 	& \qw	              & \qw & \targ{ } 	& \ctrl{-1} & \qw	            & \qw 	 & \rstick{}
	\end{quantikz}
}

\newcommand{\attackAPCtwo}{
	\begin{quantikz} [row sep= \rowSep, column sep= \colSepT, transparent]
		\lstick{$\ket{1}$} & \qw & \gate{H} & \ctrl{1} 	& \gate{R\textsubscript{Y}(\pi)} 	& \ctrl{1} 						 & \qw	\slice{n=1}   & \qw & \targ{  } & \ctrl{1} 								& \gate{R\textsubscript{Z}(\pi)} 	& \qw	\slice{n = 2} & \qw & \rstick{} \\
		\lstick{$\ket{x}$} & \qw & \gate{H} & \targ{ } 	& \ctrl{-1} 						& \gate{R\textsubscript{Z}(\pi)} & \qw	              & \qw & \ctrl{-1} & \gate{R\textsubscript{Y}(\pi)} 	    & \ctrl{-1} 						& \qw	              & \qw	& \rstick{}
	\end{quantikz}
}

\def\BibTeX{{\rm B\kern-.05em{\sc i\kern-.025em b}\kern-.08em
    T\kern-.1667em\lower.7ex\hbox{E}\kern-.125emX}}
\begin{document}

\setlength{\abovedisplayskip}{3pt}
\setlength{\belowdisplayskip}{3pt}
\setlength{\abovedisplayshortskip}{3pt}
\setlength{\belowdisplayshortskip}{3pt}

\title{Crosstalk Attack Resilient RNS Quantum Addition
		
\thanks{This research used resources of the Oak Ridge Leadership Computing Facility, which is a DOE Office of Science User Facility supported under Contract DE-AC05-00OR22725.}
}

\author{\IEEEauthorblockN{Bhaskar Gaur} 
\IEEEauthorblockA{\textit{University of Tennessee} \\
Knoxville, TN, USA \\
bgaur@vols.utk.edu}
\and
\IEEEauthorblockN{Himanshu Thapliyal} 
\IEEEauthorblockA{\textit{University of Tennessee} \\
	Knoxville, TN, USA \\
	hthapliyal@utk.edu}
}

\maketitle

\begin{abstract}
 As quantum computers scale, the rise of multi-user and cloud-based quantum platforms can lead to new security challenges. Attacks within shared execution environments become increasingly feasible due to the crosstalk noise that, in combination with quantum computer's hardware specifications, can be exploited in form of crosstalk attack. Our work pursues crosstalk attack implementation in ion-trap quantum computers. We propose three novel quantum crosstalk attacks designed for ion trap qubits: (i) Alternate CNOT attack (ii) Superposition Alternate CNOT (SAC) attack (iii) Alternate Phase Change (APC) attack. We demonstrate the effectiveness of proposed attacks by conducting noise-based simulations on a commercial 20-qubit ion-trap quantum computer. The proposed attacks achieve an impressive reduction of up to 42.2\% in output probability for Quantum Full Adders (QFA) having 6 to 9-qubit output. Finally, we investigate the possibility of mitigating crosstalk attacks by using Residue Number System (RNS) based Parallel Quantum Addition (PQA). We determine that PQA achieves higher attack resilience against crosstalk attacks in the form of 24.3\% to 133.5\% improvement in output probability against existing Non Parallel Quantum Addition (NPQA). Through our systematic methodology, we demonstrate how quantum properties such as superposition and phase transition can lead to crosstalk attacks and also how parallel quantum computing can help secure against these attacks.
\end{abstract}

\begin{IEEEkeywords}
Quantum circuit, quantum computing, quantum adder, security, trapped-ion
\end{IEEEkeywords}
\vspace*{-2mm}
\section{Introduction}

Quantum computers promise considerable advantages over classical computing, but their susceptibility to various attacks, such as fault injection, privacy, and scheduling, poses significant security challenges \cite{saki2021survey, chen2024nisq, xu2023classification}. Crosstalk noise is an inherent challenge in quantum computing, arising from unwanted interactions between qubits during multi-qubit operations \cite{ash2020analysis}. In multi-tenant quantum environments, where multiple users share mutually disjoint sets of qubits but execution starts at the same time, crosstalk can enable malicious actors to interfere with neighboring quantum circuits, leading to faulty computations, data leakage, or program disruption \cite{harper2024crosstalk, maurya2024understanding}. Therefore, studying and mitigating crosstalk attacks is crucial to ensuring the reliability of quantum computing systems.

These are the most important contributions of our work:
\begin{itemize} [leftmargin=+.3cm, topsep=2pt]
	\item We propose three crosstalk attacks and establish their feasibility in multi-tenant scenario using noise-based emulator of a commercial 20-qubit ion-trap quantum computer.
	\item We prove the effectiveness of proposed attacks by 42.2\% drop in output probability for existing Non Parallel Quantum Addition (NPQA) of 6 to 9-qubit output size.
	\item We demonstrate the attack resilience of Residue Number System (RNS) based Parallel Quantum Addition (PQA) in form of output probability improvement over NPQA ranging from 24.3\% to 133.5\%.
\end{itemize}
\vspace*{-4mm}

\begin{figure}[H]
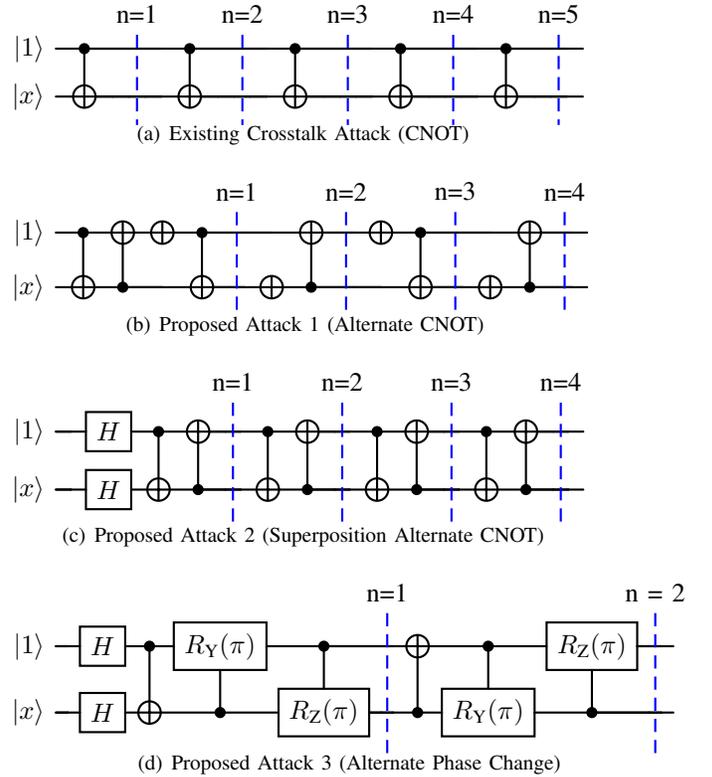

	\begin{subfigure}[Existing Crosstalk Attack (CNOT)]
		\attackCNOTtwo
	\end{subfigure}
	\begin{subfigure}[Proposed Attack 1 (Alternate CNOT)]
		\attackAltCNOT
	\end{subfigure}
	\begin{subfigure}[Proposed Attack 2 (Superposition Alternate CNOT)]
		\attackSACtwo
	\end{subfigure}
	\begin{adjustbox}{left}
		\parbox{0.2\textwidth}{
		\begin{subfigure}[Proposed Attack 3 (Alternate Phase Change)]
			\attackAPCtwo
		\end{subfigure}
	}\end{adjustbox}
	\caption{Quantum circuits for crosstalk attacks, where $\ket{x}$ is input state and n$\geq$1 is layer count expandable with Victim circuit's depth.}
	\label{fig:fig_attacks}
\end{figure}
\vspace*{-4mm}
As shown in Figure \ref{fig:fig_attacks}(a), existing crosstalk attack consists of a continuous array of CNOT gates, with the control qubit and a target qubit, initialized $\ket{1}$ and $\ket{x}$, respectively. Equation \ref{equation:CNOT} shows target qubit switching between states $\ket{x}$ and $\ket{\overline{x}}$ generating crosstalk noise. For the maximum impact, the target qubit is purposefully located next to one of the victim's qubits, making Attack 1  more suitable for superconducting quantum computers having physical connectivity \cite{harper2024crosstalk}.
\vspace*{-0.5mm}
\begin{equation} \label{equation:CNOT}
\ket{\psi\textsubscript{Existing Attack}} = \ket{1} \otimes \ket{(\sim)^nx}, \text{where }\;n \geq 1
\end{equation}

\section{Proposed Crosstalk Attacks}
\label{ProposedAttacks}

In this section, we propose three novel crosstalk attacks effective in targeting ion-trap quantum computers with any-to-any connectivity.

\subsection{Alternate CNOT (Proposed Attack 1)}
Unlike existing attack having switching activity limited to one qubit, the proposed Alternate CNOT attack, referred as Attack 1, uses CNOT gates alternatively on both qubits for control/input. Figure \ref{fig:fig_attacks}(b) displays how NOT gates separate the CNOT gates helping convert intermittent output $\ket{0}$ into $\ket{1}$, generating input for the next CNOT gate. As shown in Equation \ref{equation:AltCNOT}, Attack 1 leads to switching activity on both qubits, increasing the chances of crosstalk in ion-trap quantum computers as now any one of the attacking qubits can cause crosstalk. Here, $\ket{x}$ is input state and n is the layer count.
\begin{equation} \label{equation:AltCNOT}
	\ket{\psi\textsubscript{Attack1}} = \ket{(\sim)^{(n-1)}1} \otimes \ket{(\sim)^{(n-1)}x}, \text{where }\;n \geq 1
\end{equation}

\subsection{Superposition Alternate CNOT (Proposed Attack 2)}
While previously discussed attacks only utilized basis states, Superposition Alternate CNOT (SAC) attack or Attack 2, uses Hadamard gates on both qubits to introduce superposition. As evident from Figure \ref{fig:fig_attacks}(c), Attack 2 does not need NOT gates like Attack 1. Equation \ref{equation:SAC} shows the four mixed states that alternate among the phases 0 and \text{$\pi$}, introducing switching activity in form of phase differences across the four states.
\vspace*{-0.25mm}
\begin{multline} \label{equation:SAC}
	\ket{\psi\textsubscript{Attack2}} = \frac{1}{2}\ket{00} +
	\frac{(-1)^{(n+1+\lfloor\frac{n+2}{3}\rfloor)}}{2}\ket{01} + \\
	\frac{(-1)^{(n+\lfloor\frac{n+1}{3}\rfloor)}}{2}\ket{10} + 
	\frac{(-1)^{(n-1+\lfloor\frac{n}{3}\rfloor)}}{2}\ket{11}
	, \text{for }\;n \geq 1
\end{multline}

\subsection{Alternate Phase Change (Proposed Attack 3)}
Our final proposed Alternate Phase Change attack, referred to as Attack 3, utilizes a continuous series of flipped CNOT, Controlled-Y (CY), and Controlled-Z (CZ) gates shown in Figure \ref{fig:fig_attacks}(d) . Attack 3 causes the four mixed states to alternate between phases 0, \text{$\pi$/2}, \text{$\pi$}, and \text{3$\pi$/2}, leading to entanglement entropy alternating between 4.805e-16 and 2.73e-15, ultimately introducing more noise in the system \cite{adami1997neumann}. Qubits in ion-trap quantum computers exist in form of ions driven in a shared trap for gate operations \cite{bruzewicz2019trapped}. Higher entropy of attack qubits, if transferred to victim qubits due to crosstalk or leakage of control signal, can increase the effectiveness of the crosstalk attack \cite{saki2021shuttle}. Equations \ref{equation:APC1} and \ref{equation:APC2} represent the four states for odd (1, 3, 5..) and even integers (2, 4, 6..) respectively. 
\vspace*{-0.25mm}
\begin{multline} \label{equation:APC1}
	\ket{\psi\textsubscript{Attack3}} = \frac{\ket{00}}{2}	+\frac{(-i)^{k+1}}{2}\ket{01} -\frac{i}{2}(-1)^k\ket{10} +\frac{(-i)^k}{2}\ket{11}\\
	 \text{where k represents odd integer series (2n+1), }\;n \geq 0
\end{multline}

\begin{multline} \label{equation:APC2}
	\ket{\psi\textsubscript{Attack3}} = \frac{\ket{00}}{2}	+\frac{(-1)^{j+1}}{2}\ket{01} -\frac{(i)^j}{2}\ket{10} -\frac{(-i)^j}{2}\ket{11}\\
	\text{where j represents even integer series (2n), }\;n \geq 1
\end{multline}

\section{RNS based Parallel Quantum Addition}
\label{RNS}

Residue Number System (RNS) is an encoding scheme that represents integers based on their remainders (residues) when divided by a set of co-prime moduli \cite{mohan2016residue}. RNS allows parallel, carry-free arithmetic operations such as addition, subtraction, and multiplication, improving computational efficiency in specific applications like cryptography, signal processing, and fault-tolerant computing \cite{kalmykov2022error, roetteler2017quantum, ceschini2023modular}. In quantum computing, RNS allows Parallel Quantum Addition (PQA) across multiple jobs or quantum computers. Existing QFA of different architectures (ripple carry, carry look-ahead etc.)  have carry dependency resulting in Non Parallel Quantum Addition (NPQA) \cite{cuccaro2004new, draper2004logarithmic, takahashi2005linear, thapliyal2013design}. Compared to NPQA, RNS based PQA is noise-resilient and capable of scaling quantum addition beyond the qubit limits \cite{gaur2024residue}.
\vspace*{-4mm}
\begin{figure}[h]
	\centering
	\includegraphics[scale=0.27]{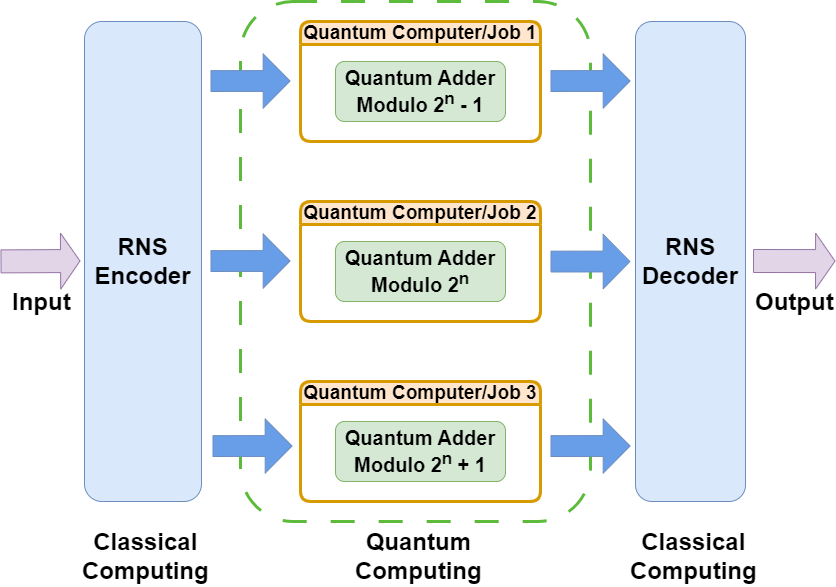}
	\caption{Residue Number System (RNS) based Parallel Quantum Addition (PQA) involves moduli set selection and Quantum Modulo Adders (QMA) generation during encoding, that can be executed in parallel without carry dependency across multiple quantum computers or jobs \cite{gaur2024residue}.}
	\label{fig:qsmart}
\end{figure}
\vspace*{-7mm}

\section{Attack Model}
\label{Methodology}
Ion trap machines have low crosstalk, allow any-to-any qubit connectivity, and can purportedly allow multiple users simultaneously (multi-tenant) as they scale up. This cloud-based usage model can help reduce cost overheads for users executing quantum circuits using fewer qubits. The attacker shares the ion-trap quantum computer in a multi-tenant scenario, with two models of attack: 1) \underline{Black Box Model}: The attacker has no access to details of the victim's quantum circuit and ends up dispatching CNOT chains of multiple widths and maximum length allowed as per the quantum volume of the quantum computer. 2) \underline{Grey Box Model}: The attacker has access to the victim circuit's width and depth, and uses that to customize the attack vector having depth slightly higher than the victim's circuit and a limited width that can help co-locate with Victim. For instance, a 6-qubit output QFA (victim) utilizes eleven qubits, leaving nine free qubits on the 20-qubit commercial ion trap quantum machine. An 8-qubit attack vector comprising four CNOT chains can co-locate in a multi-tenant scenario. Also, measurements are performed only on victim and not on attacker to avoid additional error source.

\section{Results}
\label{Results}
In this section, we conduct crosstalk-based attacks to demonstrate the attack resilience of Parallel Quantum Addition (PQA) over Non Parallel Quantum Addition (NPQA) by utilizing an emulator of a commercial 20-qubit ion-trap quantum computer. For NPQA, we conduct simulations using Quantum Full Adder (QFA) proposed by Thapliyal et al. due to its lower gate count and higher noise-resilience over other quantum adders \cite{bgaur2023noise, thapliyal2013design}. We perform simulations in 6-qubit to 9-qubit output range. For PQA, we conduct simulations for Quantum Modulo Adders (QMA) ranging from modulo 3 to modulo 9 output. These QMA can then be utilized to create Residue Number System (RNS) based parallel quantum addition for moduli set (2\textsuperscript{n}-1, 2\textsuperscript{n}, 2\textsuperscript{n}+1) \cite{kim2021quantum, gaur2024residue}. These circuits are initialized in \text{$|0$}⟩ and \text{$|1$}⟩ basis states based on the input. We calculate the average of three input cases for each adder to determine the output probability. The first case is when both inputs A and B are zero. The second case is when both A and B are equal to the maximum allowed input, which is (2\textsuperscript{n-1}-1) for a QFA with n-bit output and (k-1) for modulo k QMA. Finally, the third case is when A equals zero, and B receives the maximum allowed input. This strategy ensures fairness while evaluating adders of different types, such as QMA with QFA.
\vspace*{-2mm}
\subsection{Effectiveness of Crosstalk Attacks on NPQA}
\begin{figure}[h]
	\centering
	\includegraphics[scale=0.3]{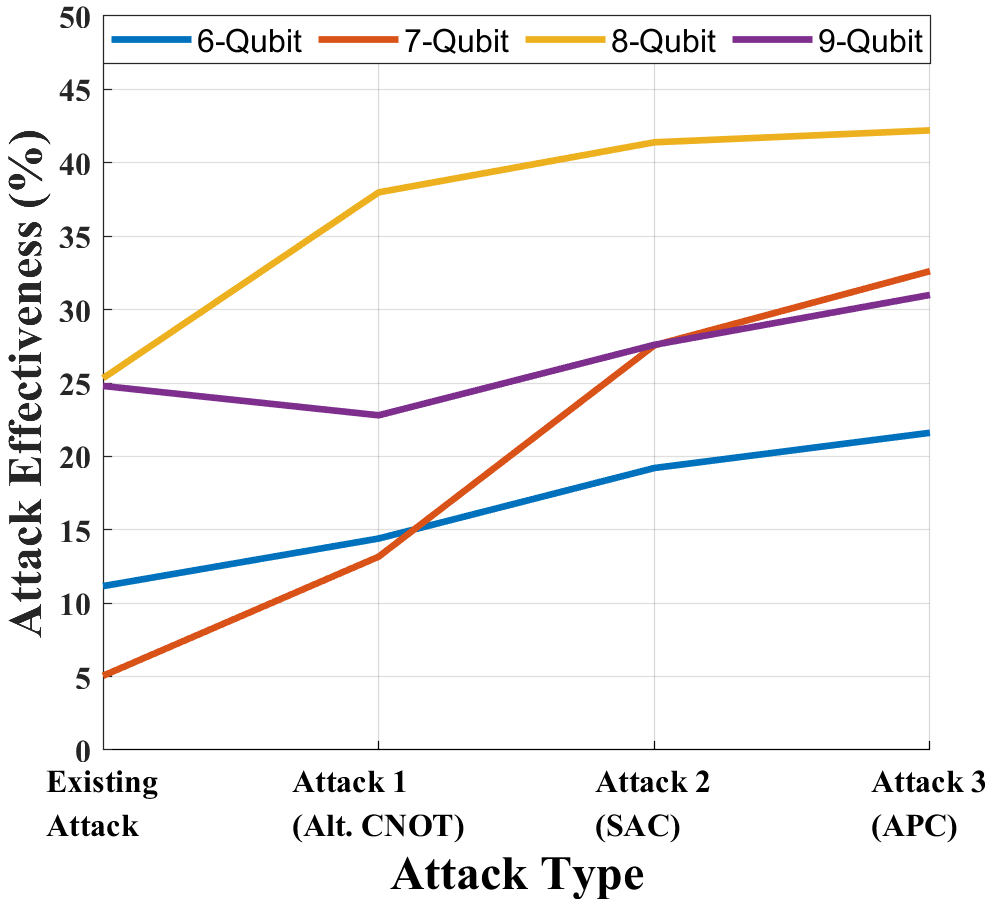}
	\vspace*{-1mm}
	\caption{Attack effectiveness (\%), calculated using output probability reduction of existing and proposed attacks over Non Parallel Quantum Addition (NPQA) for 6 to 9-qubit size.}
	\label{fig:attack_eff}
\end{figure}

Table \ref{table:comparison} shows the resource usage comparison while Table \ref{table:comparison2} shows the output probability for TPL13-based Non Parallel Quantum Addition for output size ranging from 6 to 9 qubits. While it is evident that with increasing adder size from 6-qubit to 9-qubit output, the CNOT and Toffoli depth increases, and the output probability without attack reduces from 0.833 to 0.5, a drop of about 40\%. The same trend is also evident for all attacks. For existing attack, the drop is 49.19\%, while for attack 1, this decline is 45.86\%. For attacks 2 and 3, the decrease in output probability is 46.21\% and 47.17\% respectively. This trend shows the effectiveness of crosstalk attacks as the depth of victim circuit increases. Figure \ref{fig:attack_eff} shows the effectiveness of crosstalk attacks in form of reduction in output probability when the attacks are performed against NPQA on a per adder basis from 6-qubit to 9-qubit output size. For example, a drop of 10\% in output probability would represent a 10\% increase in attack effectiveness. For the same adder size, Figure \ref{fig:attack_eff} shows the incremental effectiveness of attacks for each adder size. For instance, 7-qubit adder shows effectiveness of 5.054\% due to existing attack.
\vspace*{-1mm}
\begin{table}[!htbp]
	\centering
	\caption{Resource Usage Comparison of NPQA with PQA.}
	\label{table:comparison}
	\setlength{\tabcolsep}{3pt}
	\renewcommand{\arraystretch}{1.2}
	\begin{tabular}{|c|ccc|cccc|}
		\hline
		\multirow{3}{*}{\textbf{\begin{tabular}[c]{@{}c@{}}Adder\\ Size\end{tabular}}} & \multicolumn{3}{c|}{\textbf{\begin{tabular}[c]{@{}c@{}}Non Parallel Quantum\\ Addition (NPQA)\end{tabular}}} & \multicolumn{4}{c|}{\textbf{\begin{tabular}[c]{@{}c@{}}Parallel Quantum\\ Addition (PQA)\end{tabular}}} \\ \cline{2-8} 
		& \multicolumn{1}{c|}{\multirow{2}{*}{\textbf{\begin{tabular}[c]{@{}c@{}}Qubit\\ Count\end{tabular}}}} & \multicolumn{2}{c|}{\textbf{Depth}} & \multicolumn{1}{c|}{\multirow{2}{*}{\textbf{RNS Set}}} & \multicolumn{1}{c|}{\multirow{2}{*}{\textbf{\begin{tabular}[c]{@{}c@{}}Max.\\ Qubits\end{tabular}}}} & \multicolumn{2}{c|}{\textbf{Max. Depth}} \\ \cline{3-4} \cline{7-8} 
		& \multicolumn{1}{c|}{} & \multicolumn{1}{c|}{\textbf{Toffoli}} & \textbf{CNOT} & \multicolumn{1}{c|}{} & \multicolumn{1}{c|}{} & \multicolumn{1}{c|}{\textbf{Toffoli}} & \textbf{CNOT} \\ \hline
		\textbf{6} & \multicolumn{1}{c|}{11} & \multicolumn{1}{c|}{9} & 13 & \multicolumn{1}{c|}{(3, 4, 5)} & \multicolumn{1}{c|}{11} & \multicolumn{1}{c|}{6} & 5 \\ \hline
		\textbf{7} & \multicolumn{1}{c|}{13} & \multicolumn{1}{c|}{11} & 16 & \multicolumn{1}{c|}{(4, 5, 9)} & \multicolumn{1}{c|}{14} & \multicolumn{1}{c|}{9} & 7 \\ \hline
		\textbf{8} & \multicolumn{1}{c|}{15} & \multicolumn{1}{c|}{13} & 19 & \multicolumn{1}{c|}{(5, 8, 9)} & \multicolumn{1}{c|}{14} & \multicolumn{1}{c|}{9} & 7 \\ \hline
		\textbf{9} & \multicolumn{1}{c|}{17} & \multicolumn{1}{c|}{15} & 22 & \multicolumn{1}{c|}{(7, 8, 9)} & \multicolumn{1}{c|}{14} & \multicolumn{1}{c|}{12} & 10 \\ \hline
	\end{tabular}
\end{table}

However, Attack 1 demonstrates effectiveness of 13.17\%, an incremental increase of 8.116\% over existing attack. Similarly Attack 2 and Attack 3 have 27.57\% and 32.62\% higher effectiveness respectively demonstrating incremental effectiveness of each attack. However, the same is not evident for adder 9-qubit adder, where the attack effectiveness remains range bound between 22.8\% and 31\%. The reason seems to be the higher width of 9-qubit adder at seventeen qubits, leaving only two qubits for a single attack vector. Whereas, the 7-qubit adder utilizes thirteen qubits leaving six qubits for three attack vectors, leading to more variation. However, higher depth of 9-qubit NPQA provides opportunity for higher effectiveness regardless of attack type.

\vspace*{-3mm}
\vspace*{-4mm}

\subsection{Attack Resilient RNS based Parallel Quantum Addition}
Residue Number System (RNS) encodes integers by calculating residues using a set of co-prime moduli. To this end, we use the Quantum Modulo Adders (QMA) from (2\textsuperscript{n}-1, 2\textsuperscript{n}, 2\textsuperscript{n}+1) to select candidates for constructing the RNS moduli set. Table \ref{table:moduli} displays the resource usage, depth, and output probability of QMA from modulo 3 to modulo 9 output. The trend of decreasing output probability with increasing circuit depth is also observed in QMA, regardless of attack type. Modulo 2\textsuperscript{n} adders are least affected by attacks as they are resource efficient compared to other moduli. The moduli adder with the highest Toffoli depth, modulo 7 of type (2\textsuperscript{n}-1) with a depth of 12 Toffoli gates, also has the lowest output probability of 0.893. Modulo 9 of type (2\textsuperscript{n}+1) with a depth of 9 Toffoli follows with an output probability of 0.911. However, when simulated under attacks, the performance of modulo 7 is better than modulo 9. Modulo 7 output probability declines maximum up to 0.84 or 5.9\% due to attack 4, while the output probability of modulo 9 drops by 12.2\% from 0.911 to 0.8 due to attack 4. Modulo 9 adder requires a 4-qubit output compared to modulo 7 adder's 3-qubit output, needing an extra measurement that increases susceptibility to attacks. 

\begin{table*}[h]
	\centering
	\caption{Comparison of Quantum Moduli Adders (QMA) based on Quantum Resource Usage and Output Probability.}
	\label{table:moduli}
	\setlength{\tabcolsep}{4pt}
	\renewcommand{\arraystretch}{1.2}
	\begin{tabular}{|cl|c|cc|cc|c|c|c|c|c|}
		\hline
		\multicolumn{2}{|c|}{\textbf{Modulo Adder}} & \multirow{2}{*}{\textbf{Qubit}} & \multicolumn{2}{c|}{\textbf{Depth}} & \multicolumn{2}{c|}{\textbf{Count}} & \multirow{2}{*}{\textbf{\begin{tabular}[c]{@{}c@{}}Base Case\\ (No Attack)\end{tabular}}} & \multirow{2}{*}{\textbf{\begin{tabular}[c]{@{}c@{}}Existing\\ Attack\end{tabular}}} & \multirow{2}{*}{\textbf{\begin{tabular}[c]{@{}c@{}}Attack 1\\ (Alt. CNOT)\end{tabular}}} & \multirow{2}{*}{\textbf{\begin{tabular}[c]{@{}c@{}}Attack 2\\ (SAC)\end{tabular}}} & \multirow{2}{*}{\textbf{\begin{tabular}[c]{@{}c@{}}Attack 3\\ (APC)\end{tabular}}} \\ \cline{1-2} \cline{4-7}
		\multicolumn{1}{|c|}{\textbf{Mod}} & \multicolumn{1}{c|}{\textbf{Type}} &  & \multicolumn{1}{c|}{\textbf{Toffoli}} & \textbf{CNOT} & \multicolumn{1}{c|}{\textbf{Toffoli}} & \textbf{CNOT} &  &  &  &  &  \\ \hline
		\multicolumn{1}{|c|}{\textbf{3}} & 2\textsuperscript{n}-1 & 7 & \multicolumn{1}{c|}{6} & 7 & \multicolumn{1}{c|}{8} & 8 & 0.967 & 0.94 & 0.94 & 0.927 & 0.933 \\ \hline
		\multicolumn{1}{|c|}{\textbf{3}} & 2\textsuperscript{n}+1 & 8 & \multicolumn{1}{c|}{4} & 2 & \multicolumn{1}{c|}{5} & 2 & 0.978 & 0.956 & 0.944 & 0.956 & 0.956 \\ \hline
		\multicolumn{1}{|c|}{\textbf{4}} & 2\textsuperscript{n} & 4 & \multicolumn{1}{c|}{1} & 1 & \multicolumn{1}{c|}{1} & 2 & 0.989 & 0.989 & 0.967 & 0.978 & 0.978 \\ \hline
		\multicolumn{1}{|c|}{\textbf{5}} & 2\textsuperscript{n}+1 & 11 & \multicolumn{1}{c|}{6} & 5 & \multicolumn{1}{c|}{8} & 7 & 0.94 & 0.92 & 0.927 & 0.927 & 0.911 \\ \hline
		\multicolumn{1}{|c|}{\textbf{7}} & 2\textsuperscript{n}-1 & 10 & \multicolumn{1}{c|}{12} & 10 & \multicolumn{1}{c|}{14} & 12 & 0.893 & 0.887 & 0.873 & 0.867 & 0.84 \\ \hline
		\multicolumn{1}{|c|}{\textbf{8}} & 2\textsuperscript{n} & 6 & \multicolumn{1}{c|}{3} & 4 & \multicolumn{1}{c|}{3} & 6 & 0.978 & 0.955 & 0.944 & 0.944 & 0.922 \\ \hline
		\multicolumn{1}{|c|}{\textbf{9}} & 2\textsuperscript{n}+1 & 14 & \multicolumn{1}{c|}{9} & 7 & \multicolumn{1}{c|}{11} & 13 & 0.911 & 0.878 & 0.822 & 0.811 & 0.8 \\ \hline
	\end{tabular}
\end{table*}
\vspace*{-2mm}

\vspace*{-2mm}
\begin{figure}[h]
	\centering
	\includegraphics[scale=0.3]{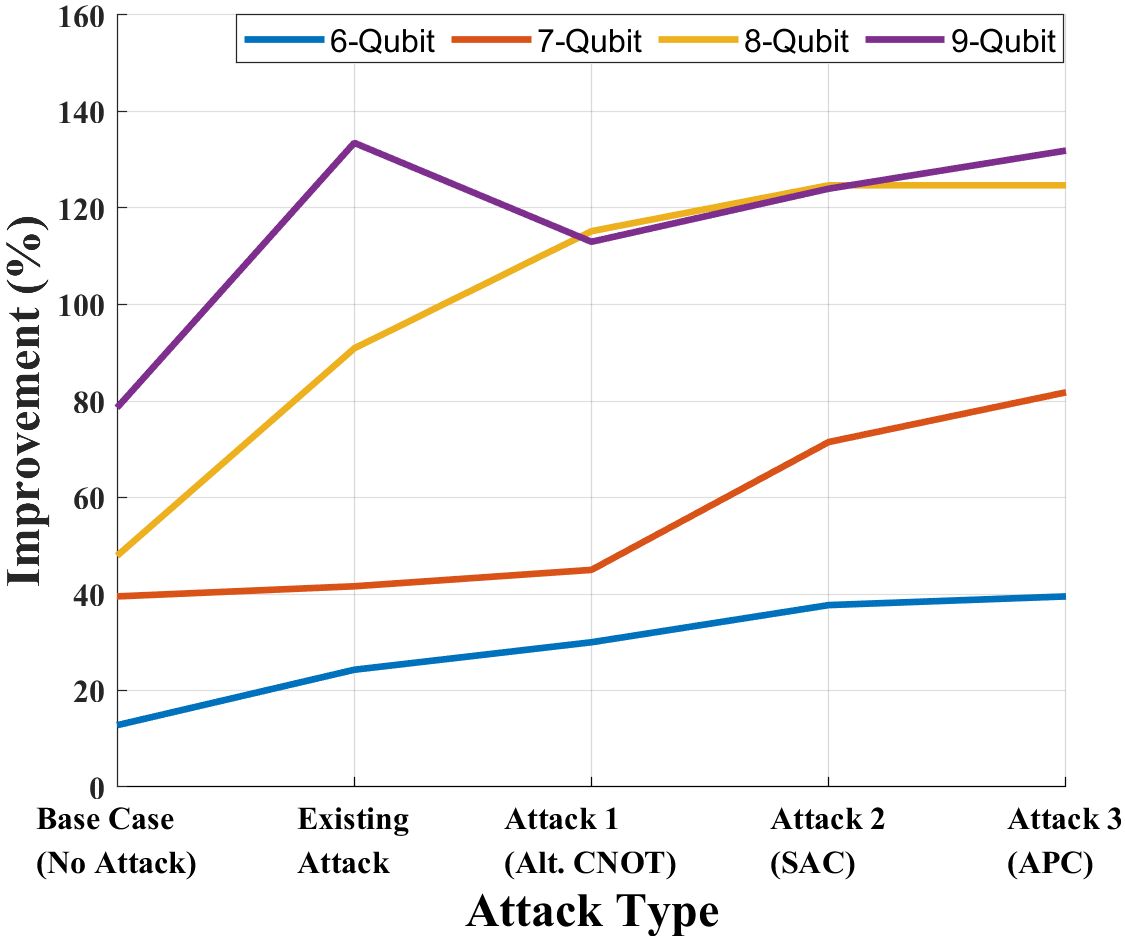}
	\vspace*{-4mm}
	\caption{Improvement (\%) Comparison of Output Probability between Non Parallel Quantum Addition (NPQA) and Parallel Quantum Addition (PQA) for output sizes 6 to 9 qubits without attack (base case), and ranging over existing and proposed crosstalk attacks.}
	\label{fig:impr_percent}
\end{figure}
\vspace*{-2mm}

\vspace*{-2mm}
\begin{table*}[h]
	\centering
	\caption{Output Probability Comparison of Non-Parallel Quantum Addition (NPQA) with Parallel Quantum Addition (PQA).}
	\label{table:comparison2}
	\setlength{\tabcolsep}{3.5pt}
	\renewcommand{\arraystretch}{1.25}
	\begin{tabular}{|c|ccc|ccc|ccc|ccc|ccc|}
		\hline
		\multirow{2}{*}{\textbf{\begin{tabular}[c]{@{}c@{}}Adder\\ Size\end{tabular}}} & \multicolumn{3}{c|}{\textbf{\begin{tabular}[c]{@{}c@{}}Base Case\\ (No Attack)\end{tabular}}} & \multicolumn{3}{c|}{\textbf{\begin{tabular}[c]{@{}c@{}}Existing Attack\\ (CNOT)\end{tabular}}} & \multicolumn{3}{c|}{\textbf{\begin{tabular}[c]{@{}c@{}}Proposed Attack 1\\ (Alternate CNOT)\end{tabular}}} & \multicolumn{3}{c|}{\textbf{\begin{tabular}[c]{@{}c@{}}Proposed Attack 2\\ (SAC)\end{tabular}}} & \multicolumn{3}{c|}{\textbf{\begin{tabular}[c]{@{}c@{}}Proposed Attack 3\\ (APC)\end{tabular}}} \\ \cline{2-16} 
		& \multicolumn{1}{c|}{\textbf{NPQA}} & \multicolumn{1}{l|}{\textbf{PQA}} & \textbf{\begin{tabular}[c]{@{}c@{}}Impr. w.r.t\\ NPQA (\%)\end{tabular}} & \multicolumn{1}{c|}{\textbf{NPQA}} & \multicolumn{1}{c|}{\textbf{PQA}} & \textbf{\begin{tabular}[c]{@{}c@{}}Impr. w.r.t\\ NPQA (\%)\end{tabular}} & \multicolumn{1}{c|}{\textbf{NPQA}} & \multicolumn{1}{c|}{\textbf{PQA}} & \textbf{\begin{tabular}[c]{@{}c@{}}Impr. w.r.t\\ NPQA (\%)\end{tabular}} & \multicolumn{1}{c|}{\textbf{NPQA}} & \multicolumn{1}{c|}{\textbf{PQA}} & \textbf{\begin{tabular}[c]{@{}c@{}}Impr. w.r.t\\ NPQA (\%)\end{tabular}} & \multicolumn{1}{c|}{\textbf{NPQA}} & \multicolumn{1}{c|}{\textbf{PQA}} & \textbf{\begin{tabular}[c]{@{}c@{}}Impr. w.r.t\\ NPQA (\%)\end{tabular}} \\ \hline
		\textbf{6} & \multicolumn{1}{c|}{0.833} & \multicolumn{1}{c|}{0.94} & 12.8 & \multicolumn{1}{c|}{0.74} & \multicolumn{1}{c|}{0.92} & 24.3 & \multicolumn{1}{c|}{0.713} & \multicolumn{1}{c|}{0.927} & 30 & \multicolumn{1}{c|}{0.673} & \multicolumn{1}{c|}{0.927} & 37.7 & \multicolumn{1}{c|}{0.653} & \multicolumn{1}{c|}{0.911} & 39.5 \\ \hline
		\textbf{7} & \multicolumn{1}{c|}{0.653} & \multicolumn{1}{c|}{0.911} & 39.5 & \multicolumn{1}{c|}{0.62} & \multicolumn{1}{c|}{0.878} & 41.6 & \multicolumn{1}{c|}{0.567} & \multicolumn{1}{c|}{0.822} & 45 & \multicolumn{1}{c|}{0.473} & \multicolumn{1}{c|}{0.811} & 71.5 & \multicolumn{1}{c|}{0.44} & \multicolumn{1}{c|}{0.8} & 81.8 \\ \hline
		\textbf{8} & \multicolumn{1}{c|}{0.616} & \multicolumn{1}{c|}{0.911} & 47.9 & \multicolumn{1}{c|}{0.46} & \multicolumn{1}{c|}{0.878} & 90.9 & \multicolumn{1}{c|}{0.382} & \multicolumn{1}{c|}{0.822} & 115.2 & \multicolumn{1}{c|}{0.361} & \multicolumn{1}{c|}{0.811} & 124.7 & \multicolumn{1}{c|}{0.356} & \multicolumn{1}{c|}{0.8} & 124.7 \\ \hline
		\textbf{9} & \multicolumn{1}{c|}{0.5} & \multicolumn{1}{c|}{0.893} & 78.6 & \multicolumn{1}{c|}{0.376} & \multicolumn{1}{c|}{0.878} & 133.5 & \multicolumn{1}{c|}{0.386} & \multicolumn{1}{c|}{0.822} & 113 & \multicolumn{1}{c|}{0.362} & \multicolumn{1}{c|}{0.811} & 124 & \multicolumn{1}{c|}{0.345} & \multicolumn{1}{c|}{0.8} & 131.9 \\ \hline
	\end{tabular}
\end{table*}
The Table \ref{table:comparison2} demonstrates attack resilience in form of improvement in output probability of Residue Number System (RNS) based Parallel Quantum Addition (PQA) over Non Parallel Quantum Addition (NPQA) for sizes 6-qubit to 9-qubit output. As evident from Figure \ref{fig:impr_percent}, the attack resilience of PQA increases as the adder size increases for all attacks. While for 6-qubit addition, the improvement from existing attack to attack 3 ranges from 24.3\% to 39.5\% respectively. For 8-qubit addition, improvement against same attacks ranges from 90.9\% to 124.7\%. This effectively demonstrates that as the size of quantum addition increases, the effect of attacks saturate in PQA as it is more depth efficient with fewer measurements per circuit. Also, for 9-qubit addition we notice a range bound improvement in attack resilience between 113\% to 133.5\%, as the effect of attacks saturate when victim occupies seventeen qubits on the commercial 20-qubit ion-trap quantum computer.

\section{Conclusion}
\label{conclusion}
In this work, we determine that crosstalk attacks are a severe problem for ion-trap quantum computers. We discover that the use of superposition and phase changes can distribute switching activity more effectively across the CNOT chain, resulting in more effective crosstalk attacks. We propose three crosstalk attacks that are proven to reduce the output probability of Non Parallel Quantum Addition (NPQA) up to 42.2\% using noise based simulations of a commercial 20-qubit ion-trap quantum computer. Finally, we show the attack resilience of Residue Number System (RNS) based Parallel Quantum Addition (PQA) over NPQA by showing improvement in output probability in the range of 24.3\% and 133.5\% when performing crosstalk attacks in a multi-tenant setting. Our findings not only expose security vulnerability in quantum full addition but also illustrate the potential of parallel quantum computing as a countermeasure against these attacks. Our work significantly enhances the understanding of crosstalk attacks and their mitigation, paving the way for more secure quantum computing systems.

\bibliographystyle{IEEEtran}
\bibliography{IEEEabrv, references}

\end{document}